\documentstyle[12pt]{ioplppt}

\def\bbt#1{\bibitem{#1} \label{bb:#1}}

\def\bsigma{\mbox{\boldmath $\sigma$}}
\def\bxi{\mbox{\boldmath $\xi$}}

\def\E{\mbox{\rm E}}
\def\Var{\mbox{\rm Var}}
\def\ustr#1#2{\;\,\stackrel{#1}{#2}\;\,}

\let\Journal=\it

\def\el#1{{\Journal Europhys. Lett.} {\bf #1}}
\def\jpa#1{{\Journal J. Phys. A: Math. Gen.} {\bf #1}}
\def\jpc#1{{\Journal J. Phys. C} {\bf #1}}
\def\jpp#1{{\Journal J. Physique} {\bf #1}}
\def\jsp#1{{\Journal J. Stat. Phys.} {\bf #1}}
\def\nn#1{{\Journal Neural Networks} {\bf #1}}

\def\pra#1{{\Journal Phys. Rev. A}{\bf #1}}
\def\pre#1{{\Journal Phys. Rev. E} {\bf #1}}
\def\prl#1{{\Journal Phys. Rev. Lett.} {\bf #1}}

\begin{document}
\title{$Q$-Ising neural network dynamics~: \\a comparative review of
        various architectures}
\author{
D.~Boll\'e\ftnote{1}{E-mail: desire.bolle@fys.kuleuven.ac.be}, 
G.~Jongen\ftnote{3}{E-mail: greetje.jongen@fys.kuleuven.ac.be}
and
G.~M.~Shim\ftnote{2}{E-mail: gmshim@nsphys.chungnam.ac.kr}}
\address{\dag\S\ Instituut voor Theoretische Fysica,
            K.U.\ Leuven, B-3001 Leuven, Belgium }
\address{\ddag\ Department of Physics, Chungnam National
            University \\Yuseong, Taejon 305-764, R.O.~Korea
}

\begin{abstract}
\noindent
This contribution reviews the parallel dynamics of $Q$-Ising neural
networks for various architectures: extremely diluted asymmetric,
layered feedforward, extremely diluted symmetric, and fully connected.
Using a probabilistic signal-to-noise ratio analysis, taking into
account all feedback correlations, which are strongly dependent
upon these architectures the evolution of the distribution of the local
field is found. This leads to a recursive scheme determining the
complete time evolution of the order parameters of the network.
Arbitrary $Q$ and mainly zero temperature are considered.
For the asymmetrically diluted and the layered feedforward
network a closed-form solution is obtained while for the symmetrically
diluted and fully connected architecture the feedback correlations
prevent such a closed-form solution.
For these symmetric networks equilibrium
fixed-point equations can be derived under certain conditions on the
noise in the system. They are the same as those obtained in a
thermodynamic replica-symmetric mean-field theory approach.
\end{abstract}
\date{}

\section{Introduction}

Artificial neural networks have been widely applied to memorize and
retrieve information. During the last few years there has been
considerable interest in neural networks with multistate neurons
(see \cite{BJSF} and references cited therein)
in order to function as associative memories for gray-toned or coloured
patterns or to allow for a more complicated internal structure in the
retrieval, e.g., a distinction between background and patterns.

Here we review the dynamics of so-called $Q$-Ising neural networks
(see the references in \cite{BVZ}) for arbitrary $Q$. They are built
{}from $Q$-Ising spin-glasses \cite{SK,GS} with couplings
defined in terms of patterns through a learning rule. For $Q=2$
one finds back the Hopfield model \cite{Hopfielda,Hopfieldb}, for
$Q \rightarrow \infty$ one has an analog network (see \cite{KB} and
references therein). One of the aims of these networks is to memorize
a number of patterns and find them back as attractors of the retrieval
process. Consequently these networks are also interesting {}from the
point of view of dynamical systems.

Besides a learning rule one also needs to specify an architecture
indicating how the spins (=neurons) are connected with each other.
Several architectures have been studied in the literature for different
purposes. {}From a practical point of view mostly perceptrons or, more
general, feedforward layered networks are used since a very long time
(see, e.g., \cite{HKP} for a history).
Hopfield \cite{Hopfielda,Hopfieldb} studied a fully connected network with
symmetric couplings because it satisfies the detailed balance principle
and hence a Hamiltonian can be defined.  Asymmetrically diluted models
\cite{DGZ} were used because their dynamics can be solved exactly  and
because they can learn us something about the loss of information
content  when some of the synaptic couplings break down.

In this contribution we review the study of the parallel dynamics of
these types of network using a probabilistic approach
(see, e.g., \cite{PZFC1,PZFC2}).
In more detail, employing a signal-to-noise ratio analysis based on the
law of large numbers (LLN) and the central limit theorem (CLT) we derive
the evolution of the distribution of the local field at every time step.
This allows us to obtain a recursive scheme for the evolution of the
relevant order parameters in the system being, in general, the main overlap
for the condensed pattern, the mean of the neuron activities and the
variance of the residual overlap responsible for the intrinsic noise in
the dynamics of the main overlap (sometimes called the width parameter).
The details of this approach depend in an essential way on the
architecture because different temporal correlations are possible.

For extremely diluted asymmetric and layered feedforward architectures
recursion relations have been obtained in closed form directly for the
relevant order parameters \cite{DGZ,BSVZ,DKM,BSV}. This
has been possible because in these types of networks there
are no feedback correlations as time progresses. As a technical
consequence the local field contains only Gaussian noise leading to an
explicit solution.

For the parallel dynamics of networks with symmetric connections,
however, things are quite different \cite{BJSF,PZFC1,PZFC2}. Even for
extremely diluted  versions of these systems \cite{BJSsymdil,WS,PZSD}
feedback correlations become essential {}from the second time step onwards,
complicating  the dynamics in a nontrivial way.
Therefore, explicit results concerning the time evolution of the order
parameters for these models have to be obtained indirectly by starting
{}from  the distribution of the local field.
Technically speaking, both for the symmetrically diluted and fully
connected  architectures the local field contains both a discrete and a
normally distributed part. The difference between the diluted and
fully connected models is that the discrete part at a certain time $t$
does  not involve the
spins at all previous times $t-1, t-2, \ldots$ up to $0$ but only the
spins at time step $t-1$. But in both cases the discrete part prevents
a closed-form solution of the dynamics for the relevant order
parameters. Nevertheless,  the development of a recursive scheme is
possible in order to calculate their complete time evolution.
In this way a comparative  discussion of the parallel dynamics at zero
temperature for the various architectures specified above is possible.

Finally, by requiring the local field to become time-independent implying
that some correlations between its Gaussian and discrete noise parts are
neglected we can obtain
fixed-point equations for the order parameters. It turns out that they
are equivalent to the fixed-point equations obtained through a
thermodynamic replica-symmetric mean-field theory approach.

At this point we remark that we do not aim for complete rigour in our
derivations. From the point of view of rigorous mathematics, the
Hopfield model and, in general, spin-glass theory is recognized to be
an extremely difficult, if not imposible, field. For a recent overview
of the modest results obtained, mostly concerning thermodynamics, we
refer to \cite{BG}.

The rest of this contribution is organized as follows. In Section
\ref{sec:mod} we
introduce the model, its dynamics and the Hamming distance as a macroscopic
measure for the retrieval quality. In Section \ref{sec:gensch} we use the
probabilistic approach in order to derive a recursive scheme for the
evolution of the distribution of the local field, leading to recursion
relations for the order parameters. The differences between the
various architectures are outlined. We do not aim for complete rigour
and mostly concentrate on zero temperature.
In Section \ref{sec:fixp} we discuss the evolution of
the system to fixed-point attractors. Some concluding remarks are given
in Section \ref{sec:con}.

\section{$Q$-Ising neural networks}
\label{sec:mod}

Consider a neural network $\Lambda$ consisting of $N$ neurons which can take
values $\sigma_i$ {}from a discrete set
        $ {\cal S} = \lbrace -1 = s_1 < s_2 < \ldots < s_Q
                = +1 \rbrace $.
The $p$ patterns to be stored in this network are supposed to
be a collection of independent and identically distributed random
variables (i.i.d.r.v.), $\{{\xi}_i^\mu \in {\cal S}\}$,
$\mu \in {\cal P}=\{1,\ldots,p\}$ and   $i \in \Lambda$,
with zero mean, $E[\xi_i^\mu]=0$, and variance $A=\Var[\xi_i^\mu]$. The
latter is a measure for the activity of the patterns. We remark that for
simplicity we have taken the patterns and the neurons out of the same
set of variables but this is no essential restriction.
Given the configuration
        ${\bsigma}_\Lambda(t)\equiv\{\sigma_j(t)\},
        j\in\Lambda=\{1,\ldots,N\}$,
the local field in neuron $i$ equals
\begin{equation}
        \label{eq:h}
        h_i({\bsigma}_{\Lambda}(t))=
                \sum_{j\in\Lambda} J_{ij}(t)\sigma_j(t)
\end{equation}
with $J_{ij}$ the synaptic coupling from neuron $j$ to neuron $i$.
In the sequel we write the shorthand notation $h_{\Lambda,i}(t) \equiv
h_i({\bsigma}_{\Lambda}(t))$.

It is clear that the $J_{ij}$ explicitly depend on the architecture.
For the extremely diluted (ED), both symmetric (SED) and asymmetric
(AED), and
the fully connected (FC) architectures the  couplings are
time-independent and the diagonal terms are absent, i.e. $J_{ii}=0$. The
configuration  ${\bsigma}_{\Lambda}(t=0)$ is chosen as input.
For the layered feedforward (LF) model the time dependence of the
couplings is relevant because the set-up of the model is somewhat
different. There
each neuron in layer $t$ is unidirectionally connected to all neurons on
layer $t+1$ and $J_{ij}(t)$ is the strength of the coupling {}{}from neuron
$j$ on layer $t$ to neuron $i$ on layer $t+1$. The state
${\bsigma}_{\Lambda}(t+1)$ of layer $t+1$ is determined by the state
${\bsigma}_{\Lambda}(t)$ of the previous layer $t$.

In all cases the couplings are chosen according to the Hebb rule such
that we can write
\begin{eqnarray}
     J_{ij}^{ED}&=&\frac{c_{ij}}{CA}
               \sum_{\mu \in {\cal P}} \xi_i^\mu \xi_j^\mu
        \quad \mbox{for} \quad i \not=j       \,,
        \label{eq:JED}  \\
     J_{ij}^{FC}&=&\frac{1}{NA}
               \sum_{\mu \in {\cal P}} \xi_i^\mu \xi_j^\mu
        \quad \mbox{for} \quad i \not=j       \,,
        \label{eq:JFC}  \\
    J_{ij}^{LF}(t)&=&\frac{1}{NA}
               \sum_{\mu \in {\cal P}} \xi_i^\mu(t+1)\, \xi_j^\mu(t)
              \,,
        \label{eq:JLF}
\end{eqnarray}
with the $\{c_{ij}=0,1\}, i,j \in \Lambda$ chosen to be i.i.d.r.v. with
distribution
$\mbox{Pr}\{c_{ij}=x\}=(1-C/N)\delta_{x,0} + (C/N) \delta_{x,1}$ and
satisfying $c_{ij}=c_{ji},\,\,\,c_{ii}=0 $ for symmetric dilution, and
$c_{ij}$ and $c_{ji}$ statistically independent (with $c_{ii}=0$) for asymmetric dilution.

At zero temperature all neurons are updated in parallel according to the
rule
\begin{equation}
        \label{eq:enpot}
        \sigma_i(t)\rightarrow\sigma_i(t+1)=s_k:
                \min_{s\in{\cal S}} \epsilon_i[s|{\bsigma}_{\Lambda}(t)]
            =\epsilon_i[s_k|{\bsigma}_{\Lambda }(t)]
\,.
\end{equation}
We remark that this rule is the zero temperature limit $T=\beta^{-1}
\rightarrow 0$ of the stochastic parallel spin-flip dynamics
defined by the transition probabilities
\begin{equation}
      \Pr \{\sigma_i(t+1) = s_k \in {\cal S} | \bsigma_{\Lambda }(t) \}
        =
        \frac
        {\exp [- \beta \epsilon_i(s_k|\bsigma_{\Lambda }(t))]}
        {\sum_{s \in {\cal S}} \exp [- \beta \epsilon_i
                                   (s|\bsigma_{\Lambda }(t))]}\,.
\label{eq:trans}
\end{equation}
Here the energy potential $\epsilon_i[s|{\bsigma}_{\Lambda}]$
is defined by \cite{R}
\begin{equation}
        \epsilon_i[s|{\bsigma}_{\Lambda}]=
                -\frac{1}{2}[h_i({\bsigma}_{\Lambda})s-bs^2]
                                            \,,
\end{equation}
where $b>0$ is the gain parameter of the system. The updating rule
(\ref{eq:enpot}) is equivalent to using a gain function $\mbox{g}_b(\cdot)$,
\begin{eqnarray}
        \label{eq:gain}
        \sigma_i(t+1) &  =   &
               \mbox{g}_b(h_{\Lambda,i}(t))
                  \nonumber      \\
               \mbox{g}_b(x) &\equiv& \sum_{k=1}^Qs_k
                        \left[\theta\left[b(s_{k+1}+s_k)-x\right]-
                              \theta\left[b(s_k+s_{k-1})-x\right]
                        \right]
\end{eqnarray}
with $s_0\equiv -\infty$ and $s_{Q+1}\equiv +\infty$. For finite $Q$,
this gain function $\mbox{g}_b(\cdot)$ is a step function.
The gain parameter $b$ controls the average slope of $\mbox{g}_b(\cdot)$.

In order to measure the retrieval quality of the system one can use the
Hamming
distance between a stored pattern and the microscopic state of the network
\begin{equation}
        d({\bxi}^\mu(t),{\bsigma}_\Lambda(t))\equiv
                \frac{1}{N}
                \sum_{i\in \Lambda}[\xi_i^\mu(t)-\sigma_i(t)]^2         \,.
\end{equation}
This  introduces the main overlap and the arithmetic mean of the
neuron activities
\begin{equation}
        \label{eq:mdef}
        m_\Lambda^\mu(t)=\frac{1}{NA}
                \sum_{i\in\Lambda}\xi_i^\mu(t)\sigma_i(t),
                \quad \mu \in {\cal P}\, ; \quad
        a_\Lambda(t)=\frac{1}{N}\sum_{i\in\Lambda}[\sigma_i(t)]^2    \,.
\end{equation}
We remark that for $Q=2$ the variance of the patterns $A=1$, and the
neuron activity $a_\Lambda(t)=1$.

\section{Solving the dynamics} \label{sec:gensch}
\subsection{Correlations}

We first discuss some of the geometric properties of the various
architectures which are particularly relevant for the understanding of
their long-time dynamic behaviour.

For a fully connected architecture there are two main sources of strong
correlations between the neurons complicating the dynamical evolution~:
feedback loops and the common
ancestor problem \cite{BKS}. Feedback loops occur when in the
course of the time evolution, e.g., the following string of connections is
possible: $i \rightarrow j \rightarrow k \rightarrow i$. We remark that
architectures with symmetric connections always have these feedback
loops. In the absence of these loops the network functions in fact as a
layered  system, i.e., only feedforward connections are possible.
But in this layered architecture common ancestors are still present when,
e.g., for  the sites $i$ and $j$ there are sites in the foregoing time
steps that have a connection with both $i$ and  $j$.

In extremely diluted asymmetric architecture these sources of
correlations are absent.
This  class of neural networks was introduced in connection with
$Q=2$-Ising  models \cite{DGZ}. We recall that the couplings are
then
given by eq.~(\ref{eq:JED}) and that in the limit $N \rightarrow \infty$
two important properties of this network are essential
\cite{DGZ,KZ}.
The first property is the high asymmetry of the connections, viz.
\begin{equation}
   \Pr\{c_{ij} = c_{ji}\} = \left(\frac{C}{N}\right)^2 \, , \quad
        \Pr\{c_{ij} = 1 \wedge c_{ji} = 0\} = \frac{C}{N}
            \left(1-\frac{C}{N}\right).
            \label{eq:G2}
\end{equation}
Therefore, the number of symmetric connections in the infinite configuration
${\bf c} = \{c_{ij}\}, i,j \neq i \in {\bf N}$ is finite with probability
one, i.e. almost all connections of the graph
$G_{{\bf N}}({\bf c}) = \{(i,j) : c_{ij}=1, i,j \neq i \in {\bf N} \}$
are  directed~: $c_{ij} \neq c_{ji}$.

The second property in the limit of extreme dilution is the directed local
Cayley-tree
structure of the graph $G_{{\bf N}}({\bf c})$. By the arguments above the
probability $F_k^{(\Lambda)}(c)$ that $k$ connections are directed towards a
given site $i \in \Lambda$ is
\begin{equation}
        F_k^{(\Lambda)}(C)
        \equiv
        \Pr \{ k = |T_i^{(in)}| \}
        =
        \frac{N!}{k!(N-k)!} \left(\frac{C}{N}\right)^k
             \left(1-\frac{C}{N}\right)^{N-k}
             \label{eq:G3}
\end{equation}
where $T_i^{(in)} = \{ c_{ji} = 1, j \in \Lambda \setminus i \}$ is the
in-tree for $i$ and $|T_i^{(in)}|$ its cardinality. This probability is
equal to
$\Pr \{ k = |T_i^{(out)}| = |\{ c_{ij} = 1, j \in \Lambda \setminus i \} | \}$
for connections directed outward a given site $i \in \Lambda$. In the limit of
extreme dilution we get a Poisson distribution :
\begin{equation}
        \lim_{N \rightarrow \infty} F_k^{(\Lambda)}(C)
        =
        \frac{C^k}{k!} e^{-C} .
\label{eq:G4}
\end{equation}
Hence, the mean value of the number of {\em in} ({\em out}) connections
for any site $i \in \Lambda$ is $E[|T_i^{(in)(out)}|] = C$. The
probability  that two
sites $i$ and $i'$ have site $j$ as a common ancestor is obviously equal to
${C}/{N}$. {}From $E[|T_i^{(in)}|] = C$ it follows that after $t$ time
steps the cardinality of the cluster of ancestors for site $i$ will be of the
order of $C^t$. The same is valid for site $i'$. Therefore, the probability
that the sites $i$ and $i'$ have disjoint clusters of ancestors
approaches $(1-{C^t}/{N})^{C^t} \simeq \exp (-{C^{2t}}/{N})$
for $N \gg 1$.

So we find that in the limit of extreme dilution :
(i)  Almost all (i.e. with probability 1) feedback loops in
$G_{{\bf N}}({\bf c})$ are eliminated.
(ii) With probability 1 any finite number of neurons have disjoint clusters
of ancestors.
So we first dilute the system by taking $N \rightarrow \infty$ and then
we take the limit $C \rightarrow \infty$ in order to get infinite
average connectivity allowing to store infinitely many patterns $p$.

This implies that for this asymmetrically diluted model at any given time
step $t$ all spins are uncorrelated and, hence, the first step dynamics
describes the full time evolution of the network.

For the symmetrically diluted model the architecture is still a local
Cayley-tree  but no longer directed and in  the limit $N \rightarrow
\infty$ the probability that the number of connections
$T_i=\{j\in \Lambda |c_{ij}=1\}$ giving information to the
the  site $i \in \Lambda$ is still a Poisson distribution with mean
$C=E[|T_i|]$. However, at time $t$ the spins are no longer uncorrelated
causing a feedback {}from $t \geq 2$ onwards \cite{PZSD}.

In order to solve the dynamics we start with a discussion of the first
time step dynamics, the form of which is independent of the architecture.

\subsection{First time step}

Consider a fully connected network. Suppose that the initial
configuration of the network
$\{\sigma_i(0)\},{i\in\Lambda}$, is a collection of i.i.d.r.v.\ with mean
$\E[\sigma_i(0)]=0$, variance $\Var[\sigma_i(0)]=a_0$, and correlated with
only one stored pattern, say the first one $\{\xi^1_i\}$:
\begin{equation}
        \label{eq:init1}
        \E[\xi_i^\mu\sigma_j(0)]=\delta_{i,j}\delta_{\mu,1}m^1_0 A
                \quad m^1_0>0 \, .
\end{equation}
This pattern is said to be condensed.  By the law of large numbers (LLN)
one  gets for the main overlap and the activity at $t=0$
\begin{eqnarray}
        m^1(0)&\equiv&\lim_{N \rightarrow \infty} m^1_\Lambda(0)
                \ustr{Pr}{=}\frac1A \E[\xi^1_i \sigma_i(0)]
                = m^1_0
                \label{eq:mo}       \\
        a(0)&\equiv&\lim_{N \rightarrow \infty} a_\Lambda (0)
                \ustr{Pr}{=} \E[\sigma_i^2(0)]=a_0
                \label{eq:a0}
\end{eqnarray}
where the convergence is in probability \cite{SH}. In order to
obtain the configuration at $t=1$ we first have to calculate the local
field (\ref{eq:h}) at $t=0$. To do this we employ the signal-to-noise
ratio analysis (see, e.g.,\cite{PZFC1,BSVZ}). Recalling the learning
rule (\ref{eq:JFC}) we separate the part containing the condensed
pattern, i.e., the signal, from the rest, i.e., the noise to arrive at
\begin{equation}
        h_i(\bsigma_{\Lambda}(0))
        =
        \xi_i^1 \frac{1}{N A} \sum_{j \in \Lambda \setminus i}
                                     \xi_j^1 \sigma_j(0)
         +
        \sqrt{\alpha}
        \frac{1}{\sqrt{pA}}
        \sum_{\mu \in {\cal P} \setminus 1 }
                \xi_i^\mu
                \frac{1}{\sqrt{NA}}
                  \sum_{j \in \Lambda \setminus i}
                       \xi_j^\mu \sigma_j(0) \nonumber\\
\label{eq:F13}
\end{equation}
where $\alpha = p/N$. The properties of the initial configurations
(\ref{eq:init1})-(\ref{eq:a0}) assure us that the summation in the first term on the r.h.s of (\ref{eq:F13}) converges in the limit $N \rightarrow \infty$ to
\begin{equation}
        \lim_{N \rightarrow \infty}
                 \frac{1}{N A} \sum_{j \in \Lambda \setminus i}
                                             \xi_j^1 \sigma_j(0)
        \stackrel{{Pr}}{=}  m^1(0).
\label{eq:F14}
\end{equation}
The first term $\xi_i^1m^1(0)$ is independent of the second term on the r.h.s of (\ref{eq:F13}). This second term contains the influence of the
non-condensed patterns causing the intrinsic noise in the dynamics of
the main overlap. In view of this we define the residual overlap
\begin{equation}
    r^\mu(t) \equiv \lim_{N \rightarrow \infty} r_{\Lambda}^\mu(t)
        =\lim_{N \rightarrow \infty}
                \frac{1}{A\sqrt{N}}\sum_{j\in \Lambda}
                \xi_j^\mu\sigma_j(t)
                \quad \mu \in {\cal P}\setminus\{1\}    \,.
        \label{eq:rdef}
\end{equation}
Applying the CLT to this second term in (\ref{eq:F13}) we find
\begin{eqnarray}
   \hspace*{-0.3cm}
     \lim_{N \rightarrow \infty} \sqrt{\frac{\alpha}{p}}
           \sum_{\mu \in {\cal P} \setminus 1}
               \xi_i^\mu r_{\Lambda \setminus i}^\mu(0)
        &=& \lim_{N \rightarrow \infty} \sqrt{\alpha}
                \frac{1}{\sqrt{p}}
                \sum_{\mu \in {\cal P} \setminus 1}
                        \xi_i^\mu
                        \frac{1}{A\sqrt{N}}
                        \sum_{j \in \Lambda \setminus i}
                                \xi_j^\mu \sigma_j(0) \\
           &\stackrel{{\cal D}}{=}&
                 \sqrt{\alpha}~{\cal N}(0,AD(0))
\label{eq:F15}
\end{eqnarray}
where the quantity ${\cal N}(0,V)$ represents a Gaussian random variable
with mean $0$ and variance $V$ and where $D(0)=\Var[r^\mu(0)]=a(0)$
Thus we see that in fact the variance of this residual overlap, i.e.,
$D(t)$ is the relevant quantity characterising the intrinsic noise.

In conclusion, in the limit $N \rightarrow \infty$ the local field is the
sum of two independent random variables, i.e.
\begin{equation}
        h_i(0)
        \equiv
        \lim_{N \rightarrow \infty} h_{{\Lambda},i}(0)
        \stackrel{{\cal D}}{=}
        \xi_i^1 m^1(0) + \sqrt{\alpha} {\cal N}(0,a(0)).
\label{eq:F16}
\end{equation}
For a more rigourous discussion of the first time step for the
underlying spin-glass model we refer to \cite{P}.
At this point we note that the structure  (\ref{eq:F16}) of
the distribution of the local field at time zero -- signal plus Gaussian
noise -- is typical for all architectures discussed here because the
correlations caused by the dynamics only appear for $t \geq 1$. Some
technical details are different for the various architectures. The first
change in details that has to be made is an adaptation of the sum over
the  sites
$j$ to $\Lambda$ for the layered feedforward architecture and to $T_i$,
the part of the tree connected to neuron $i$, in the diluted
architectures. The second change is that for
the diluted architectures an additional limit $C \rightarrow \infty$
has to be taken besides the $N \rightarrow \infty$ limit. So in the
thermodynamic limit $C, N \rightarrow \infty$ all
averages will have to be taken over the treelike structure, viz.
$\frac{1}{N}\sum_{i \in \Lambda} \rightarrow \frac{1}{C} \sum_{i \in T_j}$.
Furthermore $\alpha =p/N$ has to be replaced by $\alpha =p/C$.

\subsection{Recursive dynamical scheme}

The key question is then how these quantities evolve in time under the
parallel dynamics specified before.
For a general time step we find {}from  the
LLN in the limit $C,N \rightarrow \infty$ for the main overlap
and the activity (\ref{eq:mdef})
\begin{eqnarray}
        m^1(t+1) \ustr{Pr}{=} \frac{1}{A} \langle\!\langle
                 \xi_i^1\langle\sigma_i(t+1)\rangle_{\beta}
                                  \rangle\!\rangle , \quad
        a(t+1)   \ustr{Pr}{=} \langle\!\langle
              \langle\sigma_i(t+1)\rangle_{\beta}^2
                         \rangle\!\rangle
          \label{eq:aT}
\end{eqnarray}
with the thermal average defined as
\begin{equation}
        \langle f(\sigma_i(t+1)) \rangle_{\beta}
        =
        \frac
        {\sum_{\sigma \in {\cal S}}
        f(\sigma)
        \exp[\frac{1}{2} \beta\,\sigma(h_i(t) - b\sigma)]}
        {\sum_{\sigma \in {\cal S}}
        \exp[\frac{1}{2} \beta\,\sigma(h_i(t) - b\sigma)]}
        \label{eq:thermal}
\end{equation}
where $h_i(t) \equiv \lim_{N \rightarrow \infty} h_{\Lambda,i}(t)$.
In the above $\langle\!\langle \cdot \rangle\!\rangle$
denotes the average both over the distribution of the embedded patterns
$\{\xi_i^\mu\}$ and the  initial configurations $\{\sigma_i(0)\}$. The
average over the latter is hidden in an average over the
local field through the updating rule (\ref{eq:gain}).
In the sequel we focus on zero temperature. Then, using
eq.~(\ref{eq:gain}) these formula reduce to
\begin{eqnarray}
        m^1(t+1) \ustr{Pr}{=} \frac{1}{A} \langle\!\langle
                 \xi_i^1\mbox{g}_b(h_i(t)) \rangle\!\rangle , \quad
        a(t+1)   \ustr{Pr}{=} \langle\!\langle \mbox{g}_b^2(h_i(t))
                         \rangle\!\rangle \, .
          \label{eq:a}
\end{eqnarray}

As seen already in the first time step, we have to study carefully the
influence of the non-condensed patterns causing the intrinsic noise in
the dynamics of the main overlap.
The method used to obtain these order parameters is then to calculate
the  distribution of the local field as a function of time.
In order to determine the structure of the local field we
have to concentrate on the evolution of the residual overlap. The
details of  this calculation are very technical and depend on the
precise correlations in the system and hence on the architecture of the
network as discussed before. For these technical details we refer to the
relevant literature \cite{BJSF,BVZ,BSVZ,BSV,BJSsymdil}. Here we give an
extensive discussion of the results obtained.

In general, the distribution of the local field at time $t+1$ is given by
\begin{equation}
        h_i(t+1)=\xi_i^1m^1(t+1) + {\cal N}(0,\alpha a(t+1))
           + \chi(t) [F(h_i(t)-\xi_i^1m^1(t))+B\alpha\sigma_i(t)]
              \label{eq:hrec}
\end{equation}
where $F$ and $B$ are binary coefficients given below, which depend on the specific architecture. {}From this it is clear that the local field at time $t$ consists out of a discrete part and a normally distributed part, viz.
\begin{equation}
        h_i(t)=M_i(t) + {\cal N}(0, V(t))
\end{equation}
where $M_i(t)$ satisfies the recursion relation
\begin{equation}
        M_i(t+1)=\chi(t) [F(M_i(t)-\xi_i^1m^1(t))+B\alpha\sigma_i(t)]
                         + \xi_i^1m^1(t+1)
     \label{eq:Mrec}
\end{equation}
and where $V(t)=\alpha A D(t)$ with $D(t)$ itself given by the
recursion relation
\begin{equation}
        \label{eq:Drec}
        D(t+1)=\frac{a(t+1)}{A}+L\chi^2(t)D(t)+
                2 F\chi(t) {Cov}[\tilde r^\mu(t),r^\mu(t)] 
 \label{eq:f4}
\end{equation}
where $L$ is again a coefficient specified below. The quantity $\chi (t)$ reads
\begin{equation}
        \chi(t) = \sum_{k=1}^{Q-1} f_{\hat h_i^\mu (t)}(b(s_{k+1}+s_k))
                   (s_{k+1}-s_k) 
            \label{eq:chi}
\end{equation}
where $f_{\hat h_i^\mu (t)}$ is the probability density of  $ \hat h_i^\mu (t) = \lim _{N \to \infty} \hat h_{\Lambda,i}^\mu(t)$ with     
\begin{equation}
        \hat h_{\Lambda,i}^\mu(t)=h_{\Lambda,i}(t)-
                \frac{1}{\sqrt{N}}\xi_i^\mu r_\Lambda^\mu(t) \, . 
        \label{eq:f3}
\end{equation}
Furthermore, $\tilde r^\mu(t)$ is defined as
\begin{equation}
        \tilde r^\mu(t) \equiv \lim_{N \rightarrow \infty}
           \frac1{A\sqrt{N}}\sum_{i\in \Lambda} \xi_i^\mu
                \mbox{g}_b(\hat h_{\Lambda , i}^\mu(t)) \, .
           \label{eq:w}
\end{equation}
Finally, as can be read off {}from eq.~(\ref{eq:Mrec}) the quantity
$M_i(t)$  consists out of the signal term and a discrete noise
term, viz.
\begin{equation}
        M_i(t)=\xi _i^1 m^1(t) + B_1\alpha \chi(t-1)\sigma _i(t-1)
        + B_2\sum_{t'=0}^{t-2} \alpha
         \left[\prod_{s=t'}^{t-1} \chi(s)\right] \, \sigma _i(t')  \,.
         \label{eq:MM}
\end{equation}
Since different architectures contain different correlations not all
terms in these final equations are present. In particular we have
for the coefficients $F,B,L,B_1$ and $B_2$ introduced above
\begin{equation}
\begin{array}{l|ccccc}
&F&B&L&B_1&B_2\\
\hline
FC& 1&1&1&1&1\\
SED&0&1&0&1&0\\
LF& 0&0&1&0&0\\
AED&0&0&0&0&0
\end{array}
\end{equation}
with $B$ indicating the feedback caused by the symmetry in the
architectures and $L$ the common ancestors contribution.

At this point we remark that in the so-called theory of statistical
neurodynamics \cite{AM,Okada1996} one starts {}from an approximate
local field by leaving out any discrete noise (the term in
$\sigma_i(t)$). As a consequence the covariance in the recursion relation
for $D(t)$ can be written down more explicitly since only Gaussian noise
is involved. For more details we refer to \cite{jongen}.

We still have to determine the probability density of
$f_{\hat h_i^\mu(t)}$ in eq.~(\ref{eq:chi}), which in the thermodynamic
limit equals the probability density of $f_{h_i(t)}$. This can
be done by looking at the form of $M_i(t)$ given by eq.~(\ref{eq:MM}).
The evolution equation tells us that $\sigma _i(t')$ can be replaced by
$g_b(h_i(t'-1))$ such that the second  and third terms of $M_i(t)$ are
the sums of stepfunctions of correlated variables. These are also
correlated through the dynamics with the normally distributed
part of $h_i(t)$. Therefore the local field can be considered as a
transformation of a set of correlated normally distributed variables
$x_s,\, s=0,\ldots,t-2,t$, which we choose to normalize. Defining the
correlation matrix $W = \left(\rho(s,s')\equiv \E[x_s x_{s'}] \right)$
we  arrive at the following expression for the probability density of
the  local field at time~$t$
\begin{eqnarray}
     f_{h_i(t)}(y)&=&\int\prod_{s=0}^{t-2} dx_s dx_t ~
         \delta \left(y - M_i(t)-\sqrt{\alpha A D(t)}\,x_t\right)
             \nonumber\\
             &\times& \frac{1}{\sqrt{\mbox{det}(2\pi W)}}
            ~\mbox{exp}\left(-\frac{1}{2}{\bf x} W^{-1}
            {\bf x}^T\right)
            \label{eq:fhdisfc}
\end{eqnarray}
with ${\bf x}=(x_0,\ldots x_{t-2},x_t)$. For the symmetrically diluted
case this expression simplifies to
\begin{eqnarray}
     f_{h_i(t)}(y)&=&\int\prod_{s=0}^{[t/2]} dx_{t-2s} ~
             \delta \left(y -\xi^1_i m^1(t)- \alpha \chi(t) \sigma_i(t)
              -\sqrt{\alpha a(t)}\,x_t\right) \nonumber\\
             &\times& \frac{1}{\sqrt{\mbox{det}(2\pi W)}}
            ~\mbox{exp}\left(-\frac{1}{2}{\bf x} W^{-1}
            {\bf x}^T \right)
            \label{eq:fhdisd}
\end{eqnarray}
with ${\bf x}=(\{x_s\})=(x_{t-2[t/2]},\ldots x_{t-2},x_t)$. The brackets
$[t/2]$ denote the integer part of $t/2$.

So the local field at time $t$ consists out of a
signal term, a discrete noise part and a normally distributed noise part.
Furthermore, the discrete noise and the normally distributed noise are
correlated and this prohibits us to derive a closed expression for the
overlap and activity.

Together with the eqs.~(\ref{eq:a}) for $m^1(t+1)$ and $a(t+1)$ the
results above form a recursive scheme in order to obtain the
order parameters of the system. The practical difficulty which remains is
the explicit calculation of the correlations in the network at different
time steps as present in eq.~(\ref{eq:f4}).

For AED and LF architectures this scheme leads to an explicit form for
the recursion relations for the order parameters
\begin{eqnarray}
   m^\mu(t+1)
   &=&
   \delta_{\mu,1}\frac{1}{A} \left\langle\!\left\langle\xi^1(t+1)
   \int Dz\,\mbox{g}( \xi^1(t+1)m^1(t)+\sqrt{\alpha AD(t)}\,z)
   \right\rangle\!\right\rangle\label{eq:rm} \\
   a(t+1)
   &=&
   \left\langle\!\left\langle \int Dz\,\mbox{g}^2(\xi^1(t+1)m^1(t)
    +\sqrt{\alpha AD(t)}\,z) \right\rangle\!\right\rangle
    \label{eq:ra} \\
   D(t+1)
   &=&
   \frac{a(t+1)}{A}+\frac{L}{\alpha A}\left[ \left\langle\!\left\langle
         \int Dz\,z\mbox{g}(\xi^1(t+1)m^1(t)
        +\sqrt{\alpha AD(t)}\,z)\right\rangle\!\right\rangle \right]^2
   \,. \nonumber \\ \label{eq:rD}
\end{eqnarray}
For the AED architecture $(L=0)$ the second term on the r.h.s. of (\ref{eq:rD})
coming {}from the correlations caused by the common ancestors is absent.
For the LF architecture we remark that this explicit solution requires an
independent choice of the representations of the patterns at different
layers. At finite temperatures analogous recursion relations for the AED
and LF networks can be
derived \cite{BSVZ,BSV} by introducing auxiliary thermal fields
\cite{PZparis} in order to express the stochastic dynamics within the
gain function formulation of the deterministic dynamics. Furthermore,
damage spreading \cite{DGZ,D,DM}, i.e., the evolution of two network    
configurations which are initially close in Hamming distance can be
studied \cite{BSVZ,BSV}. Finally, a complete self-control mechanism can
be built in the dynamics of these systems by introducing a
time-dependent threshold in the gain function improving, e.g., the basins of
attraction of the memorized patterns \cite{DB}$^{-}$\cite{G}.  

For explicit examples of this dynamical scheme with numerical results we
refer to \cite{BJSF,BJSsymdil}. By using the recursion relations the
first few time
steps are written out explicitly and studied numerically, e.g, for the
$Q=2$ and $Q=3$ FC and SED models with equidistant states and a uniform
distribution of  patterns. These results are compared with the
approximations studied in the literature \cite{PZFC1,PZFC2,WS,PZSD,AM,Okada1996,K}$^{-}$\cite{GSZ} by
neglecting some feedback correlations for $t \geq 2$.
In the whole retrieval region of these networks we find that the first
four or five time steps give us already a clear picture of their time
evolution.

\section{Fixed-point equations} \label{sec:fixp}

Equilibrium results for the AED and LF $Q$-Ising models are obtained
immediately by straightforwardly leaving out the time dependence in
(\ref{eq:rm})-(\ref{eq:rD})(see \cite{BSVZ},\cite{BSV}), since the
evolution  equations for the local field and the order parameters
do not change their form as time progresses (see
eqs.~(\ref{eq:rm})-(\ref{eq:rD})).  This still allows small fluctuations in
the configurations $\{\sigma_i \}$.  The difference between the
fixed-point equations for these two architectures is that
for the AED model the variance of the residual noise, $D(t)$ is
simply proportional to the activity of the neurons at time $t$ while
for  the LF model a recursion is needed.

For the SED and FC architectures, however, the evolution equations for the
order parameters do change their form by the explicit appearance of the
$\{\sigma_i(t')\}, t'=1,\ldots,t $ term.  Hence we can not use the
simple procedure above to obtain the fixed-point equations. Instead we
derive the equilibrium results of our dynamical scheme by requiring
through the recursion relations (\ref{eq:hrec}) that the distribution
of the local field becomes time-independent. This is an approximation
because fluctuations in the network configuration are no longer allowed.
In fact, it means that out of the discrete part of
this distribution, i.e., $M_i(t)$ (recall (\ref{eq:MM})), only the
$\sigma _i(t-1)$ term is kept besides, of course, the signal term.
This procedure implies that the main overlap and activity in the
fixed-point are found {}from the definitions (\ref{eq:mdef})
and not {}from leaving out the time dependence in the recursion relation
(\ref{eq:a}).

We start by eliminating the time-dependence in the evolution equations for
the local field (\ref{eq:hrec}). This leads to
\begin{equation}
        \label{eq:hfix}
        h_i=\xi_i^1m^1 + [{\bar \chi}^{ar}]^{-1} {\cal N}(0,\alpha a)
                +[{\bar \chi}^{ar}]^{-1}\alpha \chi \sigma_i
\end{equation}
with ${\bar \chi}^{ar} \equiv 1-F\chi$ being $1$ for the SED and $1-\chi$ for the FC model and
$h_i \equiv \lim_{t \rightarrow \infty} h_i(t)$.
This expression consists out of two parts: a normally distributed part
$\tilde h_i = {\cal N}(\xi_i^1m^1,\alpha a / [{\bar \chi}^{ar}]^2)$
and some discrete noise part. At this point some remarks are in order.
First, the
discrete noise coming {}from the correlations of the $\{\sigma_i(t)\}$ at
different time steps (here only the preceding time step is considered)
is inherent in the SED and FC dynamics. Second, the so-called
self-consistent signal-to-noise ratio analysis of the FC network
considered in the literature \cite{SFa,SFb} starts {}from
such a type of equation by assuming the presence of a term
proportional to the output in the local field without any reference or
argumentation based upon the  underlying dynamics of the network.

Employing this expression in the updating rule (\ref{eq:gain}) one finds
\begin{equation}
        \label{eq:sfp}
        \sigma_i=\mbox{g}_b(\tilde h_i+
                [{\bar \chi}^{ar}]^{-1} \alpha \chi \sigma_i)
\end{equation}
where $\tilde h_i={\cal N}(\xi_i^1m^1,\alpha a)$ is the normally
distributed part of eq.~(\ref{eq:hfix}).
This is a self-consistent equation in $\sigma_i$ which
in general admits more than one solution. These types of equation have been
solved in the literature in the context of thermodynamics using a
geometric Maxwell construction \cite{SFa},\cite{SFb}.
We remark that for analog networks
the geometric Maxwell construction  is not necessary: the
fixed-point equation (\ref{eq:sfp}) has only one solution.
For more technical details we refer to \cite{jongen}.

This approach leads to a unique solution
\begin{equation}
        \sigma_i=\mbox{g}_{\tilde b}(\tilde h_i), \quad
        {\tilde b} = b - [{2\bar \chi}^{ar}]^{-1}\alpha \chi
            \,.
\end{equation}
We remark that plugging this result into the local field (\ref{eq:hfix})
tells us that the latter is the sum of two Gaussians with shifted mean
(see also \cite{HO}).

Using the definition of the main overlap and activity
(\ref{eq:mdef}) in the limit $N \rightarrow \infty$
for the FC model and limit $C,N \rightarrow \infty$ for the SED model,
one finds in the fixed point
\begin{eqnarray}
        \label{eq:m1fix}
        m^1 =\left\langle\!\left\langle\xi^1\int {\cal D}
        z ~  \mbox{g}_{\tilde b}
                \left( \xi^1m^1 + \sqrt{\alpha A D}\,z
                \right)\right\rangle\!\right\rangle \,
         \\
        \label{eq:afix}
        a   =\left\langle\!\left\langle\int {\cal D}
        z ~  \mbox{g}_{\tilde b}^2
                \left( \xi^1m^1 + \sqrt{\alpha A D}\,z
                \right)\right\rangle\!\right\rangle
          \,.\end{eqnarray}
{}From (\ref{eq:Drec}) and (\ref{eq:chi}) one furthermore sees that
\begin{equation}
        \label{eq:Dfix}
        D = [{\bar \chi}^{ar}]^{-2} a/A
\end{equation}
with
\begin{equation}
        \label{eq:chifix}
        \chi=\frac1{\sqrt{\alpha A D}}
                \left\langle\!\left\langle\int {\cal D}
                z ~  z \, \mbox{g}_{\tilde b}
                        \left( \xi^1m^1 + \sqrt{\alpha A D}\,z
                        \right)\right\rangle\!\right\rangle \,.
\end{equation}
These resulting equations (\ref{eq:m1fix})-(\ref{eq:Dfix}) are the same as
the fixed-point equations derived {}from a replica-symmetric mean-field
theory treatment in \cite{WS2}$^{-}$\cite{BRS}. Their solution
leads to capacity-gain parameter phase diagrams (see, e.g.,\cite{BRS}).

\section{Concluding remarks} \label{sec:con}

An evolution equation is derived for the distribution
of the local field governing the parallel dynamics at zero temperature
of extremely diluted symmetric and asymmetric, layered feedforward and
fully connected $Q \geq 2$-Ising networks.
All feedback correlations are taken into
account. In general, this distribution
is not normally distributed but contains a discrete noise part.

Employing this evolution equation a general recursive scheme is developed
allowing one to calculate the relevant order parameters of the
system, i.e., the main overlap, the activity and the variance of the
residual noise for any time step.
For the extremely diluted asymmetric and the layered feedforward
architectures this scheme immediately leads to explicit recursion
relations for the order parameters because the discrete noise part in
the  local field is absent.
For the extremely diluted and the fully connected architectures equilibrium
fixed-point equations for the order parameters are obtained under
the  condition that the local field becomes time-independent, meaning
that some of the discrete noise is neglected.
The resulting equations are the same as those derived
{}from a replica-symmetric mean-field theory approach.

\section*{Acknowledgments}

This work has been supported in part by the Research Fund of the
K.U.Leuven (Grant OT/94/9) and the Korea Science and Engineering
Foundation through the SRC program. The authors are indebted to
S.~Amari, R.~K\"uhn, G.~Massolo, A.~Patrick and V.~Zagrebnov for
constructive discussions.
One of us (D.B.) thanks the Belgian National Fund for Scientific
Research for financial support.

\end{document}